# STM and DFT studies of $CO_2$ adsorption on Cu(100)-O surface

Steven J. Tjung[1,†], Qiang Zhang[2,†], Jacob J. Repicky[1], Simuck F. Yuk[2], Xiaowa Nie[3], Nancy M. Santagata[1], Aravind Asthagiri[2], Jay A. Gupta[1]*

[1]*Department of Physics, The Ohio State University, Columbus, OH 43210*

[2]*William G. Lowrie Department of Chemical and Biomolecular Engineering, The Ohio State University, Columbus, OH 43210*

[3] *School of Chemical Engineering, PSU-DUT Joint Center for Energy Research, State Key Laboratory of Fine Chemicals, Dalian University of Technology, Dalian 116024, P. R. China*

**Abstract**

We characterized $CO_2$ adsorption and diffusion on the missing row reconstructed ($2\sqrt{2} \times \sqrt{2}$) R45° Cu(100)-O surface using a combination of scanning tunneling microscopy (STM) and density functional theory (DFT) calculations with dispersion. We deposited $CO_2$ molecules *in situ* at 5K, which allowed us to unambiguously identify individual $CO_2$ molecules and their adsorption sites. Based on a comparison of experimental and DFT-generated STM images, we find that the $CO_2$ molecules sit in between the O atoms in the missing row reconstructed Cu(100)-O surface. The $CO_2$ molecules are easily perturbed by the STM tip under typical imaging conditions, suggesting that the molecules are weakly bound to the surface. The calculated adsorption energy, vibrational modes, and diffusion barriers of the $CO_2$ molecules also indicate weak adsorption, in qualitative agreement with the experiments. A comparison of tunneling spectroscopy and DFT-calculated density of states shows that the primary change near the Fermi level is associated with changes to the surface states with negligible contribution from the $CO_2$ molecular states.

[†] Steven J. Tjung and Qiang Zhang contributed equally to this work.

Corresponding author: jgupta@physics.osu.edu

## I. Introduction

Heterogeneous catalysis of $CO_2$ conversion to useful chemicals and/or fuels has drawn much interest with applications including hydrogenation [1], dry reforming [2], electroreduction [3–5], and photoreduction [6–9]. Many of these applications include oxide catalysis [10–12] and characterizing the nature of the $CO_2$-surface oxide interaction is an important first step to understanding catalyst activity. For example, electron transfer in $CO_2$ photoreduction is sensitive to the relative alignment of energy levels of adsorbed $CO_2$ and the surface states of specific adsorption sites of the catalyst. Copper oxides have recently gained interest as catalysts for photoelectroreduction of $CO_2$ [13,14], but these oxide phases are complex and difficult to characterize. This challenge motivates surface science studies of $CO_2$ adsorption on model single crystal surfaces [15–19]. Among surface science methods, scanning tunneling microscopy (STM) and scanning tunneling spectroscopy (STS), combined with density functional theory (DFT) are attractive approaches to characterize interactions of $CO_2$ with copper oxide surfaces at an atomic level. Prior STM studies of $CO_2$ adsorption on Ag [20–22] and $TiO_2$ [23,24] surfaces have provided insight into the role of defects and low coordination sites (e.g., step edges) in $CO_2$ activation. Similar studies on copper oxides could shed light on the interplay of photo/electroexcitation of the molecules and surface morphology.

Cu(100)-O is an ideal model surface for copper oxides as it represents the initial transition to oxidation of the Cu(100) surface [25,26]. In addition, it is a good platform to study $CO_2$ adsorption because the surface is well characterized. The structure of oxygen on the Cu(100) surface has been well investigated with various techniques in the past [27–33] where a c(2 × 2) structure was observed for low oxygen coverage and a (2√2 × √2) R45° missing row type structure has been observed for higher oxygen coverage [28,30,34]. The latter reconstruction involves the removal of one of every four Cu atomic rows, creating a periodic missing row structure. The O atoms occupy the sites along the edges of the missing row, creating O-Cu-O chains [35]. The Cu(100)-O surface has also been characterized by detailed STM and AFM studies [33]. Though structurally well characterized, there have been few studies of adsorption on this surface [36,37].

Here we present a joint STM / DFT study of $CO_2$ adsorption on the Cu(100)-O surface at the single molecule level. Atomic resolution STM images of the missing row reconstruction are in good agreement with HSE06-DFT simulated images. Tunneling spectroscopy reveals two additional unoccupied states which are reproduced in the DFT-calculated LDOS. STM imaging before and after $CO_2$ adsorption is used to unambiguously identify individual molecules. The agreement between experimental STM images and spectroscopy and DFT calculations indicates that $CO_2$ physisorbs at a bridging oxygen site in the missing row at low temperature (5K). The linear structure, vibrational frequencies and molecular orbital states of physisorbed $CO_2$ on Cu(100)-O closely resemble the gas phase. Small changes in the Cu(100)-O surface electronic structure upon $CO_2$ adsorption are revealed in tunneling spectroscopy and the differential LDOS and can be attributed to the surface states.

## II. Methods

The STM experiments were performed with a CreaTec LT-STM operating at 5K in ultrahigh vacuum (UHV) with a base pressure of ~ $7 \times 10^{-11}$ mbar. Samples were prepared in an attached UHV chamber ($10^{-10}$ mbar) before being transferred into the cold STM. A clean Cu(100) surface was first prepared with repeated cycles of $Ar^+$ sputtering and annealing at 550°C. To obtain the oxygen-induced surface reconstruction, $10^{-6}$ mbar of $O_2$ gas was introduced into the preparation chamber via a precision leak valve. The Cu(100) substrate was exposed to $O_2$ for 5 min while being held at a temperature of ~ 300°C. Auger electron spectroscopy was used to verify that the surface contamination was < 1 % of a monolayer, and to monitor the oxygen coverage. After exposure, the $O_2$ was pumped out and the sample cooled down to 100 K within 30 minutes, before being transferred into the cold STM. Once in the STM, the sample cooled down to 5 K in ~ 6 hours.

Constant current STM images were collected using a PtIr tip. Tunneling spectroscopy in constant height or constant current mode was performed using a lock-in method by adding a 5-40mV modulation at 1453 Hz to the bias voltage. The modulation in measured tunneling current provides information about the sample's local density of states (LDOS). In constant height mode, the STM feedback loop is turned off after the tip is positioned and the tunneling current is recorded as a function of the bias voltage. In the constant current mode, the STM feedback loop remains active and maintains a constant tunneling current as the bias voltage is varied. This mode allows us to obtain tunneling spectra over a larger voltage range not limited by the dynamic range of the current amplifier. STM images were analyzed using the WSxM program [38].

$CO_2$ molecules were introduced into the STM chamber through a precision leak valve at a pressure of $5 \times 10^{-9}$ mbar for 5 minutes. A wobble stick was used to open small holes in the radiation shields surrounding the STM allowing the $CO_2$ molecules to adsorb onto the sample surface at 5K. *In situ* dosing allowed us to study the adsorption site of $CO_2$ molecules on the O-Cu(100) surface by comparing STM images of the same area before and after the dose.

All DFT calculations were performed using the Vienna ab initio simulation package (VASP) [39,40] using projector-augmented-wave pseudopotentials available in VASP database [41,42]. A plane-wave cutoff of 400 eV was applied with a Fermi-smearing width of 0.2 eV. Exchange and correlation effects were described with the Perdew-Burke-Ernzerhof (PBE) form of the generalized gradient approximation (GGA) functional [43]. Because $CO_2$-surface interactions are expected to have a significant van der Waals interaction, we employed the dispersion-corrected DFT proposed by Grimme and co-workers (referred to as DFT-D3) [44,45]. The PBE lattice constant for Cu is 3.64 Å, which agrees well with the experimental value of 3.61 Å and also prior DFT studies [46], was used to fix the lateral dimension of the Cu(100) slab. The bare Cu(100) surface is modeled with a 4 layer slab with the bottom 2 layers fixed. A force criterion of 0.03 eV/Å was used for geometry optimization and a 15 Å vacuum spacing, along with dipole corrections[47], was used to minimize spurious periodic interactions along the surface normal.

We examined $CO_2$ adsorption on both (2×2) and (4×4) surface unit cells with corresponding (4×4×1) and (2×2×1) Monkhorst-Pack k-point meshes, respectively [48]. All STM and adsorption energies reported in the paper are obtained using a (4×4) surface unit cell, but initial exploration of possible $CO_2$ adsorption sites was done on the (2×2) surface unit cell. The adsorption energy is defined as Eq. (1), where $E_{CO2}$ is the isolated $CO_2$ molecule energy, $E_{slab}$ is the energy of the relaxed O chemisorbed Cu(100), and $E_{ads-slab}$ is the energy of $CO_2$ adsorbed on the slab.

$$E_{ads} = (E_{slab} + E_{CO2}) - E_{ads-slab} \quad (1)$$

With the definition in Eqn. 1, a larger positive $E_{ads}$ signifies a more stable adsorption site for $CO_2$. We determined the barriers and the pathway of $CO_2$ diffusion using the climbing nudged elastic band (cNEB) method [49].

For the surface and most stable $CO_2$ adsorption configuration, DFT-STM images were obtained using the Tersoff-Hamann approximation, where the tunneling current is proportional to the local density of state (LDOS) of the surface at the tip position integrated from the Fermi level ($E_f$) to the applied voltage bias ($E_F + eV_{bias}$) [50]. The simulated STM images were calculated from a (4×4) surface and generated by the p4vasp software package [51]. We choose a isodensity of 0.001 electrons/Å$^3$ within p4vasp since it provided the best images in comparison to experiment. Because of the sensitivity of the STM images to the surface LDOS, we performed single point (i.e. no relaxation) non-local hybrid HSE06 exchange-correlation functional [52] to examine the electronic structure of the bare O-Cu(100) surface and the most favored $CO_2$ configuration on the surface.

### III. Results and Discussion

**Cu(100)-O Structure**

The ordered phase of oxygen on Cu(100) has been well characterized with various techniques such as LEED [27], HREELS [28], XRD[29], and STM [30–33]. These studies indicate that the (2√2 x √2) R45° reconstructed surface corresponding to a 0.5 ML coverage can be achieved under UHV conditions. Top and side views of the ball model of the (2√2 x √2) R45° Cu(100)-O surface are shown in Figure 1a. This model shows alternating O and Cu atoms (O-Cu-O) rows which are separated by a row of Cu atoms on one side, and a 'missing' row of vacant sites on the other side. The oxygen atoms are nearly planar with the Cu atoms. Figure 1b shows an STM image of the Cu(100)-O surface, with a ladder-like contrast resulting from the (2√2 x √2) R45° missing row reconstruction. This result is consistent with previous STM studies.[30–33] There is remarkably little surface contamination, evidenced by the complete lack of point defects on the clean surface. Two domain orientations with 90° rotation are observed which are attributed to nucleation along the [001] and the [010] directions of the Cu(100) surface [30]. An atomically resolved STM image of the surface can be seen in Figure 1c. Here we see bright protrusions which form the rungs and rails of the ladder structure. An overlay of the model of the Cu(100)-O surface allows us to assign the bright protrusions in Figure 1c as the surface Cu atoms while the O atoms are imaged as the gaps between the rungs of the ladder structure.

For comparison, an HSE06 simulated STM image of the Cu(100)-O surface with the same bias voltage as experiment is shown in figure 1d. A clear ladder-like structure can be observed in the HSE06 STM images where the surface Cu atoms appear as bright protrusions, in agreement with the experimental STM image. Figure S1 in the supplementary information (SI) compares the PBE and HSE06 STM images. While both methods provide qualitatively similar images, the HSE06 STM images show a closer match to the experimental STM images at similar bias voltages.

A comparison of the LDOS (see Fig. S2) shows a downward shift in the O states below the Fermi level for HSE06 versus PBE. This shift results in a reduced LDOS for surface O atoms in HSE06 compared to PBE at the image bias voltage. Thus, surface O atoms image as a depression in HSE06 (consistent with experiment), but a protrusion in PBE. These results suggest that HSE06 is more accurate in capturing the relative Cu and O states on the surface versus the Fermi level. Therefore, we will focus on comparison between experimental data and HSE06 calculations in the remaining discussion.

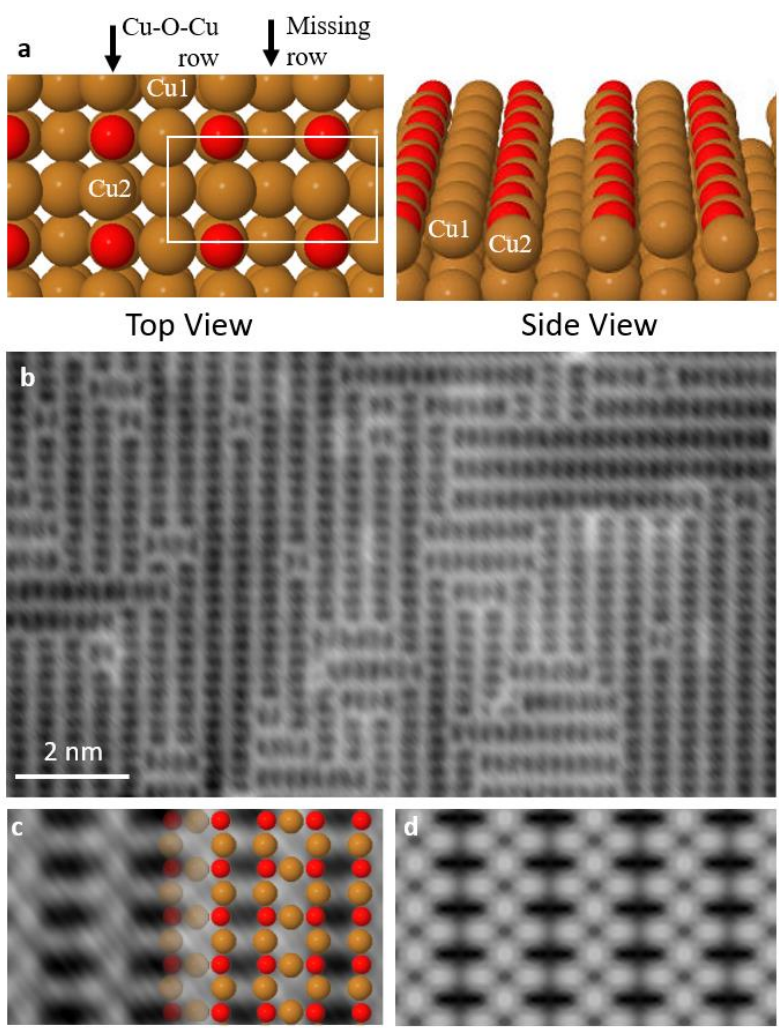

Figure 1 (a) Top and side view of O-Cu(100) surface model. Brown and red spheres represent Cu and O atoms, respectively. (b) Large area STM image of the O-Cu(100) surface, showing different domains of the missing row reconstruction. (300mV, 0.2nA). (c) Atomically resolved STM image of O-Cu(100) surface with the overlay of the model. (-400 mV, 0.2nA). (d) DFT HSE06 simulated STM image of the Cu(100)-O surface ( -400 mV, 0.001 $e/\text{Å}^3$).

While the surface structure of Cu(100)-O has been well characterized, there is only one prior photoemission study of occupied states in the surface electronic structure [53]. Here, we probed the electronic structure of Cu(100)-O by performing tunneling spectroscopy. Figure 2a compares constant height tunneling spectra from the Cu(100)-O and clean Cu(100) surfaces. Spectra from Cu(100)-O exhibit little dependence on tip position within the surface unit cell. Both sets of spectra are relatively featureless in the occupied states region (V < 0), which reflects a lack of prominent states in the voltage range probed here, as well as the fact that tunneling under these conditions is dominated by electrons near the sample Fermi level $E_{Fs}$, which makes STM less sensitive to occupied states well below $E_{Fs}$. In unoccupied states (V >0), there is a distinct rise in DOS above 1V for the Cu(100)-O surface. To compare the surface electronic structure over a wider voltage range, we also measured constant current tunneling spectra as shown in Figure 2b. The spectra in the region of occupied states (V < 0V) is again relatively featureless on both surfaces, despite the prominent Cu $d$-states which lie a few eV below $E_{Fs}$. We attribute this discrepancy to the limited STM sensitivity for low-lying occupied states. For unoccupied sample states (V > 0) of clean Cu(100), we measured a series of field emission resonances (FERs) beginning at 5.2 V due to the potential well created by the tunneling electron and its image charge [54]. This hydrogen-like series of states begins near the surface vacuum level, as determined by the surface work function. For example, the $n = 1$ FER on Cu(100) is at ~ 5.2 V, close to the surface work function of 5.1 eV [55]. On Cu(100)-O, a number of peaks are also observed and the interpretation of these peaks depends on locating where the FER series begins on this surface. For example, if the surface work function were reduced due to oxygen adsorption, one could assign the peak at 3.2 V to the $n = 1$ FER. If the work function were increased, one could assign the peaks > 5.6 V to the FER series, and the peaks at 1 and 3.2 V to additional states of the Cu(100)-O surface, similar to what we have previously observed for the Cu(100)-N surface [56].

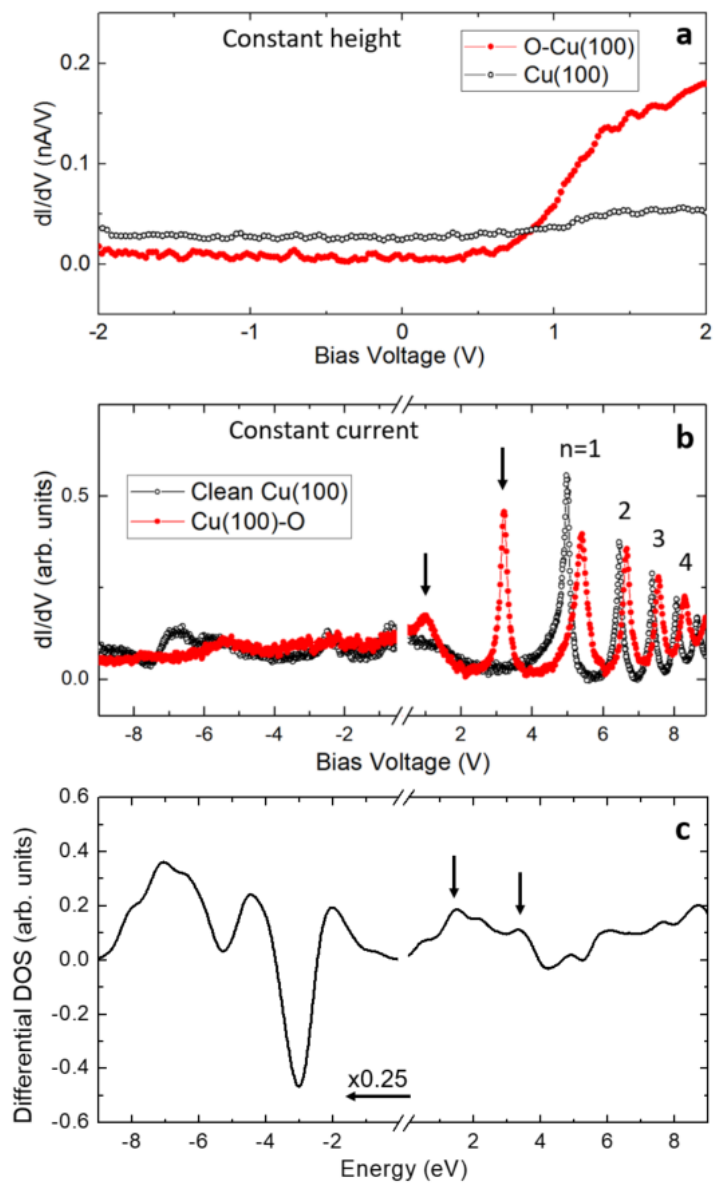

Figure 2 (a) Constant height tunneling spectra of Cu(100) and O-Cu(100) surface. The tip height was set at (0.5 V, 0.2 nA for O-Cu, and 3 V, 0.25 nA for Cu) (b) Constant current tunneling spectra of Cu(100) and Cu(100)-O. Peaks in the curves labeled as n = 1 ,2 ,… indicate FERs. Arrows indicate two additional peaks observed on Cu(100)-O. (0.2 nA for O-Cu, and 0.5 nA for Cu) (c) Calculated average HSE06 LDOS difference for surface atoms in Cu(100)-O and Cu(100). The two peaks in the calculated LDOS indicated by arrows are found at similar energies as the peaks in the experimental data.

To distinguish between these possible interpretations, we compare these experimental results with the DFT-calculated changes in work function. We find that the work function increases by 0.28 eV (PBE) and 0.50 eV (HSE06) for Cu(100)-O compared to the Cu(100) surface. This increase in work function corresponds to the surface dipole induced by the charge on the surface atoms. Prior DFT study of the Cu(100)-O surface suggested that O atoms are partially negatively charged due to electron transfer from the substrate [46]. Indeed, we find from HSE06 (PBE) a Bader charge of -1.14$e$ (-0.98$e$) on the surface O atoms and an average Bader charge of +0.62$e$ (+0.55$e$) on the first layer Cu atoms, corresponding to charge transfer from Cu to O.

The increased work function predicted from DFT leads us to assign the peak at 5.6 V in the experimental data to the $n = 1$ FER of Cu(100)-O, with the hydrogenic series extending to higher voltages. This assignment then suggests the peaks observed experimentally at 1 V and 3.2 V can be attributed to new states from the O atoms. To more clearly identify changes in electronic structure upon oxygen adsorption, we compute the difference in LDOS between Cu(100)-O (average LDOS of the surface Cu atoms + average LDOS of the surface O atoms) and Cu(100) surfaces (average LDOS of the surface Cu atoms) as shown in Figure 2c.

Focusing first on unoccupied states, the calculated differential LDOS exhibits a broad bunching of states between 1-4 eV as indicated by arrows in Figure 2c that represent an admixture of Cu and O states upon adsorption. These features are in qualitative agreement with the two peaks below 4 V observed experimentally, though a more quantitative comparison would need to consider the tunneling process and tip density of states as well. Turning now to the occupied states, DFT predicts a prominent dip at – 3 eV, attributed to contribution primarily from shifts of the Cu 3$d$ states upon O adsorption. Our calculations also show peaks at -2.0 eV, -4.5 eV, and -7.0 eV. The broad peak at -7.0 eV comes from both O states and new induced Cu states on the O-Cu(100) surface, while the other two peaks have contributions from shifted Cu states as well as smaller contribution from O states. Though states so far below $E_F$ are difficult to probe with STM spectroscopy, previous photoemission experiments did show oxygen induced features in a region of – 7 eV to – 4 eV below the Fermi level [53] in relatively good agreement with our calculations.

**CO$_2$ Adsorption on Cu(100)-O**

Next, we characterize the adsorption of CO$_2$ molecules on the Cu(100)-O surface. Figures 3a-b show STM images of the same area of the surface before and after *in situ* CO$_2$ dose at 5K. The *in situ* dosing produces new bright features, which we can unambiguously identify as individual CO$_2$ molecules. The dosing parameters we used result in a low coverage of CO$_2$ molecules. Higher CO$_2$ coverage can be obtained with longer dosing times. CO$_2$ molecules are imaged as bright protrusions with an apparent height of ~ 0.45 Å under these tunneling conditions, elongated along the ladder-like structure of the Cu(100)-O surface. The STM image also shows that CO$_2$ molecules are only adsorbed along the missing row, centered between the rungs of the ladder structure. All the CO$_2$ molecules in the STM image in Figure 3b exhibit the same contrast and shape, suggesting that there is only one orientation configuration for CO$_2$ on the Cu(100)-O surface. The absence of additional point defects with different image contrast indicates that CO$_2$ molecules are not dissociated upon adsorption, as we would expect to image adsorbed C and O atoms if that were the case.

We often observe "glitch" lines in the STM image as we scan over a CO$_2$ molecule (arrows in Fig. 3b), which we attribute to tip-induced motion of the CO$_2$ molecules during imaging. Careful line-by-line inspection of these images reveals the path of molecular motion during the glitches. The top arrow in Fig. 3b shows a molecule which twice hopped between neighboring missing rows during scanning. This produces an abrupt change in the apparent height as the tip moves. The bottom arrow indicates a molecule which also hopped twice during scanning, but this time remaining along the same missing row. The tip's influence likely extends several nm from the tunneling apex, as we occasionally observed motion of molecules between successive images without apparent glitch lines (c.f., Figure S3). We were unable to identify tunneling conditions which can completely remove the tip perturbation effect, suggesting that the $CO_2$ molecules are relatively weakly bound to the surface. To set a bound on intrinsic diffusion processes, we could retract the tip and compare images of the same area as a function of time interval (c.f., Fig. S4). In this way, we were able to confirm that no motion occurred over a period of 12 hours, indicating that intrinsic diffusion processes are slow at 5 K in the absence of tip perturbation.

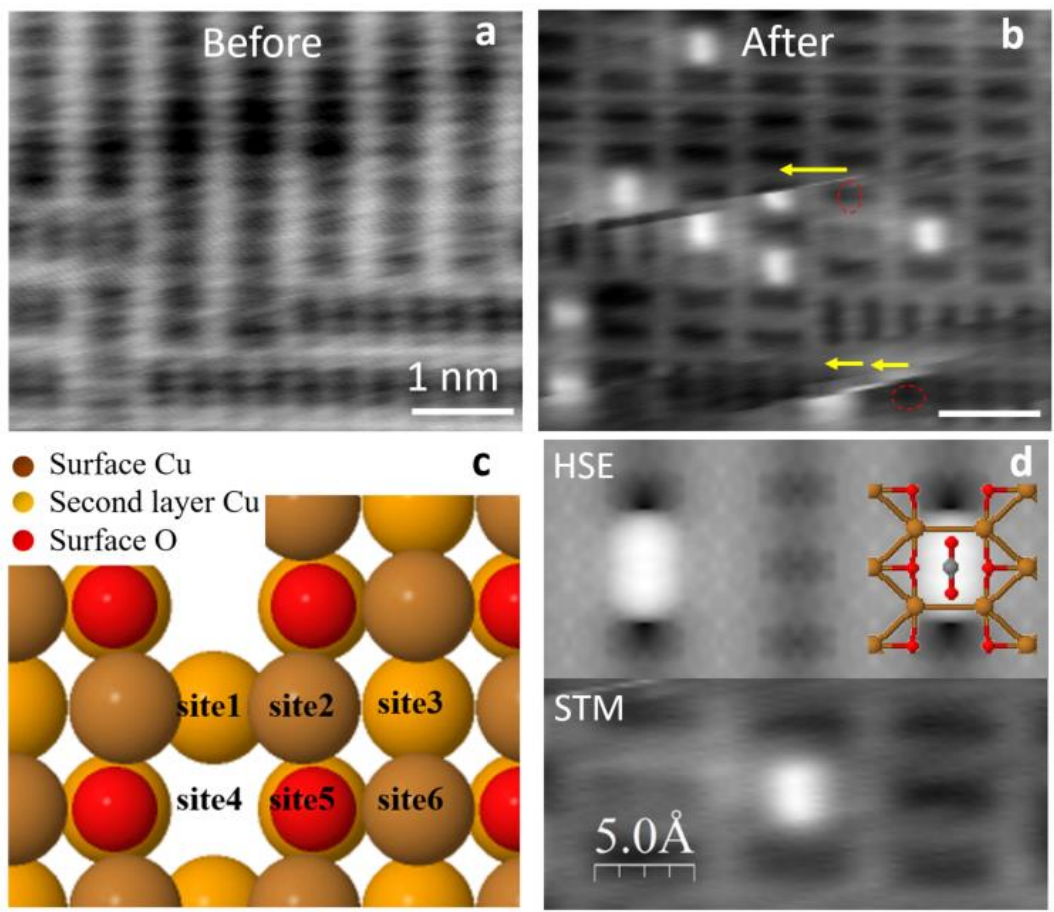

Figure 3 (a) STM image of Cu(100)-O surface before $CO_2$ dose. (b) STM image of the same area of Cu(100)-O surface after $CO_2$ dose. Glitch lines seen on the image indicate tip induced motion of the $CO_2$ molecules during imaging. The initial location of $CO_2$ before the motion is indicated by dashed red circle and the motions are indicated by the yellow arrows. (c) Model of Cu(100)-O surface showing six different adsorption sites. (d) Comparison between DFT-HSE06 simulated STM image (top) and experimental STM image (bottom) of $CO_2$ molecules adsorbed on Cu(100)-O. STM data taken at (50 mV, 0.2 nA). DFT simulation at (200 mV, 0.001 $e$/Å$^3$).

We performed HSE06-DFT calculations to gain further insight into $CO_2$ adsorption and compare with our experimental observations. Figure 3c shows six possible $CO_2$ adsorption sites on the missing row reconstructed Cu(100)-O surface examined using DFT. For each adsorption site we perform 45° rotations in the plane to give a total of 24 total configurations. In addition, for select configurations we have also examined $CO_2$ oriented perpendicular (i.e. upright) to the surface and initially slightly bent $CO_2$, but these configurations were found to be not stable. The resulting most favored $CO_2$ adsorption energies for the sites are listed in Table 1 and associated images for the relaxed structures can be found in Fig. S5 in the Supporting Information. Vibrational frequency calculations indicate that only adsorption on $Bridge_O$ (site 4) and O (site 5) sites are stable since the other four sites have imaginary vibrational modes indicating that they are transition states. Comparing the adsorption energy of the two possible stable adsorption sites, the $Bridge_O$ site is the most stable adsorption site for the $CO_2$ molecules. On this site, the $CO_2$ molecule is stabilized in a linear configuration by the surrounding four Cu atoms and two O atoms (see Fig. S5 in the SI).

We also calculated DFT-simulated STM images of the $CO_2$ molecule adsorbed on the $Bridge_O$ site as shown in figure 3d (top), along with an overlay of the DFT structural model. This confirms that the molecular axis is parallel to the missing Cu row, with the C atom adsorbed between O atoms on the surface and the O atoms in the $CO_2$ adsorbed closer to the Cu atoms in the rungs of the ladder structure. This explains the elongated contrast and adsorption site in the experimental STM images (Fig. 3d, bottom). Overall, the experimental and DFT STM images show good agreement for linear $CO_2$ to be adsorbed on the $Bridge_O$ site.

**Table 1.** PBE-D3 $CO_2$ adsorption energies (eV) on O/Cu(100). See Fig. 3c for description of adsorption sites.

|        | Site        | Adsorption energy (eV) |
|--------|-------------|------------------------|
| Site 1 | $Bridge_{Cu}$ | $0.22^{TS}$ |
| Site 2 | Cu2         | $0.23^{TS}$ |
| Site 3 | Hollow      | $0.19^{TS}$ |
| Site 4 | $Bridge_O$  | 0.32 |
| Site 5 | O           | 0.24 |
| Site 6 | Cu1         | $0.22^{TS}$ |

$^{TS.}$ The relaxed configuration has an imaginary vibrational mode, indicating a transition state.

The relatively low adsorption energy of 0.32 eV suggests weak physisorption of $CO_2$ molecules on the Cu(100)-O surface, consistent with the experimental observation that $CO_2$ can be easily perturbed during imaging. Evaluating the adsorption energy for $CO_2$ on $Bridge_O$ site with and without dispersion shows that van der Waals interactions contribute around 85% of the adsorption energy. This is not surprising, as $CO_2$ physisorbs on nearly all metals and metal oxides at low temperature (< 80K) [15]. Also supporting weak adsorption, we find small changes in $CO_2$ vibrational frequencies upon adsorption (c.f., Table S1). We also calculated diffusion barriers to better understand the motion of the molecule on the Cu(100)-O surface. Possible diffusion paths were considered using the most stable adsorption side ($Bridge_O$ site) as either initial or final state.

The images for the resulting pathways are shown in Fig. S6 in the SI. The calculated barrier for $CO_2$ to diffuse along the missing row is 0.15 eV, while the barrier for diffusion across the O-Cu-O row is 0.10 eV. These relatively small diffusion barriers, together with a low adsorption energy, are consistent with a picture of $CO_2$ physisorption on the Cu(100)-O surface. The similar diffusion barriers also indicate that we would expect motion in both directions. This is in good agreement with the experimental results, where out of 22 tip induced motions we observed, 13 were along the missing row and 9 were across the O-Cu-O rows.

We performed tunneling spectroscopy to probe changes in the Cu(100)-O surface electronic structure upon $CO_2$ adsorption. Figure 4a compares constant height dI/dV spectra taken on the $CO_2$ molecule, and on a clean area of the Cu(100)-O surface. The $CO_2$ spectra were taken with the tip positioned directly above the C atom, but we found no significant variation at other points within the molecule or on other $CO_2$ molecules, suggesting these data are representative and are limited by the spatial resolution in STM/STS. The voltage range in this measurement is limited to ± 2 V due to the instability of the $CO_2$ molecules. The $CO_2$ molecules were perturbed by the STM tip upon sweeping to higher bias voltages, leading to abrupt changes in the dI/dV signal associated with molecular motion. We find that the $CO_2$ exhibits relatively little distinct electronic structure compared to the Cu(100)-O surface, evidenced by the lack of pronounced peaks in the tunneling spectra. There is a slight increase in conductance near +/- 2 V on the $CO_2$ molecule, but no prominent states in between.

To quantitatively compare the experimental STS with DFT, we calculated the LDOS for the $CO_2$/Cu(100)-O surface. We find that the $CO_2$ contributes negligible states in this energy range. We calculated the gap between highest occupied (HOMO) and lowest unoccupied (LUMO) frontier molecular orbitals of the adsorbed $CO_2$ to be 10.58 (8.8) eV using the HSE06 (PBE) functional, compared with 9.6 (8.1) eV for an isolated $CO_2$ molecule. This change in the HOMO-LUMO gap reflects a small interaction with the surface leading to a small shift and broadening of the $CO_2$ states. Concomitantly, such a large gap results from frontier molecular orbitals that lie far from the Fermi level, outside the experimentally accessible voltage range in our tunneling spectroscopy. To better probe $CO_2$-induced changes to surface electronic structure, we compute the differential DOS by subtracting the contributions from surface atoms (Cu1, Cu2, O) with and without adsorbed $CO_2$. The differential DOS (Figure 4b, black) reflects the changes to the surface states upon $CO_2$ adsorption and would be zero if there were no differences. For comparison, we compute a differential dI/dV signal (Figure 4b, red). Considering occupied states, in Figure 4b we see the differential DOS begins to rise around -1 eV below the Fermi level, which corresponds qualitatively to the experimental data where we observe an increase of conductance at -1 V. In unoccupied states, the differential DOS is negative from 1-2 eV, in agreement with the differential dI/dV which also shows a decreased conductance in the range 1-1.75V. This discussion is consistent with a picture where the $CO_2$ does not directly contribute states near the Fermi level, but instead can induce more subtle changes in surface electronic structure that are observed experimentally.

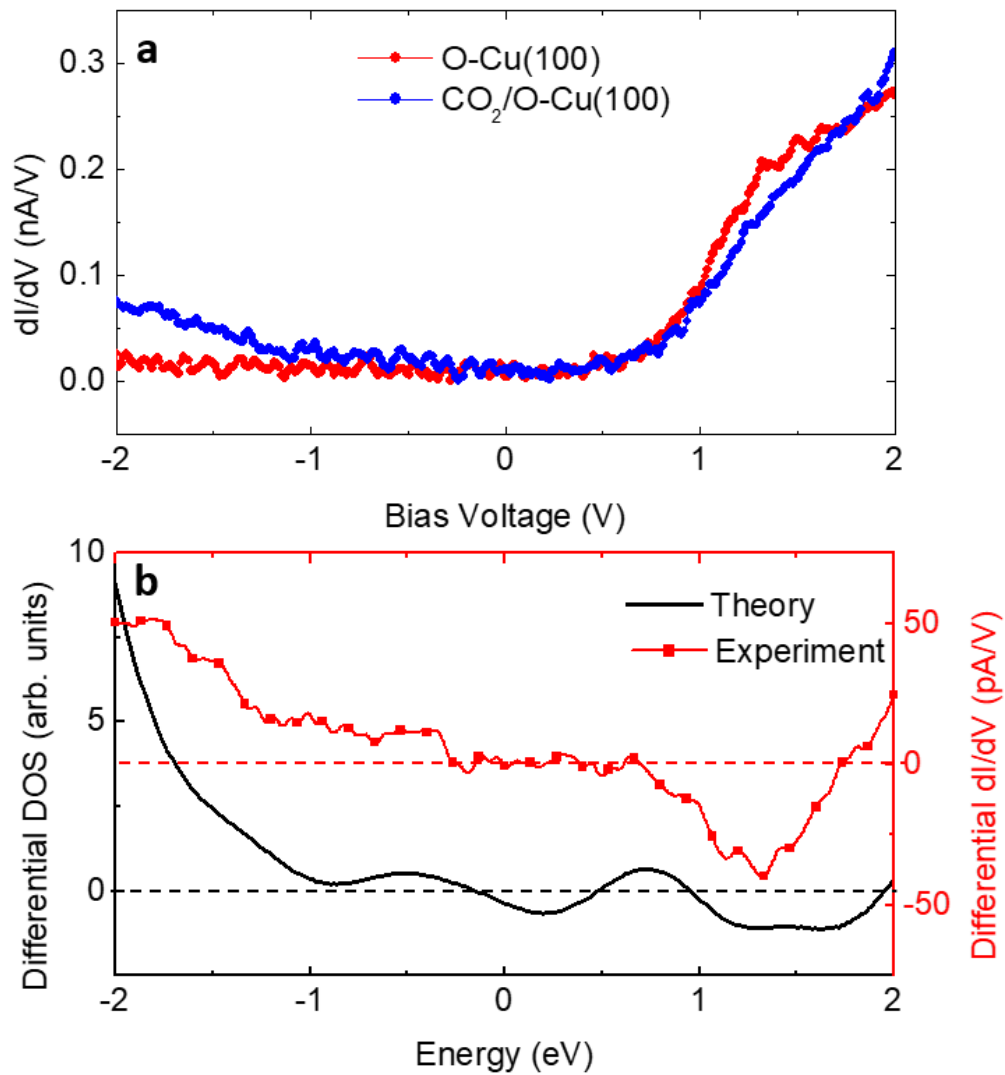

Figure 4. (a) Constant height spectroscopy of O-Cu(100) (red) and $CO_2$ molecule (blue). Tip height set at (2 V, 0.2 nA) (b) Comparison of HSE06 differential DOS {DOS[Cu1, Cu2, O for $CO_2$ adsorbed on Cu(100)-O] – DOS[Cu1, Cu2, O for bare Cu(100)-O]} and differential dI/dV signal calculated from (a). Dashed lines indicate the zeros for both data sets to distinguish positive and negative differential regions.

**Conclusion**

In conclusion, we characterized the clean Cu(100)-O surface with STM and resolved the ladder-like structure of the missing row reconstruction, consistent with previous studies. Using scanning tunneling spectroscopy, we observed two new states and a shift of the $n = 1$ FER away from the Fermi level, in good agreement with the DFT-calculated DOS and increased surface work function. *In situ* deposition of $CO_2$ allows us to unambiguously identify and study the adsorption sites of $CO_2$ on the Cu(100)-O surface. STM imaging and DFT calculations show that the most stable adsorption site for $CO_2$ is bridging between two surface O atoms. Our STM experiment and DFT calculation also show that the $CO_2$ molecules are weakly bound on the surface and have small diffusion barriers along and perpendicular to the missing rows. Tunneling spectroscopy of the $CO_2$ molecules in comparison with the DFT calculated DOS near the Fermi level shows only a relatively small change in the surface electronic structure due to $CO_2$ adsorption.

**Acknowledgement**

Funding for this research was provided by ACS PRF New Directions Grant under award number 56135-ND5 and Institute for Materials Research OSU Materials Seed Grant. We acknowledge the Ohio Supercomputing Center for providing the computational resources for this work.

**References**

[1] W.H. Wang, Y. Himeda, J.T. Muckerman, G.F. Manbeck, and E. Fujita, Chem. Rev. **115**, 12936 (2015).

[2] L. Guczi, G. Stefler, O. Geszti, I. Sajó, Z. Pászti, A. Tompos, and Z. Schay, Appl. Catal. A Gen. **375**, 236 (2010).

[3] T. Inoue, A. Fujishima, S. Konishi, and K. Honda, Nature **277**, 637 (1979).

[4] M. Gattrell, N. Gupta, and A. Co, J. Electroanal. Chem. **594**, 1 (2006).

[5] Y. Hori, in *Mod. Asp. Electrochem.*, edited by C.G. Vayenas, R.E. White, and M.E. Gamboa-Aldeco (Springer, New York, NY, 2008), pp. 89–189.

[6] J.L. White, M.F. Baruch, J.E. Pander, Y. Hu, I.C. Fortmeyer, J.E. Park, T. Zhang, K. Liao, J. Gu, Y. Yan, T.W. Shaw, E. Abelev, and A.B. Bocarsly, Chem. Rev. **115**, 12888 (2015).

[7] B. Kumar, M. Llorente, J. Froehlich, T. Dang, A. Sathrum, and C.P. Kubiak, Annu. Rev. Phys. Chem. **63**, 541 (2012).

[8] S.C. Roy, O.K. Varghese, M. Paulose, and C.A. Grimes, ACS Nano **4**, 1259 (2010).

[9] M.C. Toroker, D.K. Kanan, N. Alidoust, L.Y. Isseroff, P. Liao, and E.A. Carter, Phys. Chem. Chem. Phys. **13**, 16644 (2011).

[10] G. Yin, M. Nishikawa, Y. Nosaka, N. Srinivasan, D. Atarashi, E. Sakai, and M. Miyauchi, ACS Nano **9**, 2111 (2015).

[11] I. Shown, H.-C. Hsu, Y.-C. Chang, C.-H. Lin, P.K. Roy, A. Ganguly, C.-H. Wang, J.-K. Chang, C. Wu, L. Chen, and K.-H. Chen, Nano Lett. **14**, 6097 (2014).

[12] Q. Zhai, S. Xie, W. Fan, Q. Zhang, Y. Wang, W. Deng, and Y. Wang, Angew. Chemie - Int. Ed. **52**, 5776 (2013).

[13] G. Ghadimkhani, N.R. de Tacconi, W. Chanmanee, C. Janaky, and K. Rajeshwar, Chem. Commun. **49**, 1297 (2013).

[14] C. Janáky, D. Hursán, B. Endrődi, W. Chanmanee, D. Roy, D. Liu, N.R. de Tacconi, B.H. Dennis, and K. Rajeshwar, ACS Energy Lett. **1**, 332 (2016).

[15] H.-J. Freund and M.W. Roberts, Surf. Sci. Rep. **25**, 225 (1996).

[16] W. Taifan, J.-F. Boily, and J. Baltrusaitis, Surf. Sci. Rep. **71**, 595 (2016).

[17] U. Burghaus, Prog. Surf. Sci. **89**, 161 (2014).

[18] B. Eren, R.S. Weatherup, N. Liakakos, G.A. Somorjai, and M. Salmeron, J. Am. Chem. Soc. **138**, 8207 (2016).

[19] M. Favaro, H. Xiao, T. Cheng, W.A. Goddard, J. Yano, and E.J. Crumlin, Proc. Natl. Acad. Sci. U. S. A. **114**, 6706 (2017).

[20] I. Stensgaard, E. Laegsgaard, and F. Besenbacher, J. Chem. Phys. **103**, 9825 (1995).

[21] X.-C. Guo and R.J. Madix, Surf. Sci. **489**, 37 (2001).

[22] Y. Okawa and K. ichi Tanaka, Surf. Sci. **344**, (1995).

[23] S. Tan, Y. Zhao, J. Zhao, Z. Wang, C. Ma, A. Zhao, B. Wang, Y. Luo, J. Yang, and J. Hou, Phys. Rev. B - Condens. Matter Mater. Phys. **84**, (2011).

[24] J. Lee, D.C. Sorescu, and X. Deng, J Am Chem Soc **133**, 10066 (2011).

[25] C. Gattinoni and A. Michaelides, Surf. Sci. Rep. **70**, 424 (2015).

[26] Q. Zhu, L. Zou, G. Zhou, W.A. Saidi, and J.C. Yang, Surf. Sci. **652**, 98 (2016).

[27] H.C. Zeng, R.A. McFarlane, and K.A.R. Mitchell, Surf. Sci. **208**, L7 (1989).

[28] M. Wuttig, R. Franchy, and H. Ibach, Surf. Sci. **213**, 103 (1989).

[29] I.K. Robinson, E. Vlieg, and S. Ferrer, Phys. Rev. B **42**, 6954 (1990).

[30] F. Jensen, F. Besenbacher, E. Laegsgaard, and I. Stensgaard, Phys. Rev. B **42**, 9206 (1990).

[31] C. Wöll, R.J. Wilson, S. Chiang, H.C. Zeng, and K.A.R. Mitchell, Phys. Rev. B **42**, 11926 (1990).

[32] F.M. Leibsle, Surf. Sci. **337**, 51 (1995).

[33] M.Z. Baykara, M. Todorović, H. Mönig, T.C. Schwendemann, Ö. Ünverdi, L. Rodrigo, E.I. Altman, R. Pérez, and U.D. Schwarz, Phys. Rev. B **87**, 155414 (2013).

[34] T. Fujita, Y. Okawa, Y. Matsumoto, and K. Tanaka, Phys. Rev. B **54**, 2167 (1996).

[35] P.J. Knight, S.M. Driver, and D.P. Woodruff, J. Phys. Condens. Matter **9**, 21 (1997).

[36] H. Altass, A.F. Carley, P.R. Davies, and R.J. Davies, Surf. Sci. **650**, 177 (2016).

[37] B. Eren, H. Kersell, R.S. Weatherup, C. Heine, E.J. Crumlin, C.M. Friend, and M.B. Salmeron, J. Phys. Chem. B **122**, 548 (2018).

[38] I. Horcas, R. Fernández, J.M. Gómez-Rodríguez, J. Colchero, J. Gómez-Herrero, and A.M. Baro, Rev. Sci. Instrum. **78**, 013705 (2007).

[39] G. Kresse and J. Furthmüller, Comput. Mater. Sci. **6**, 15 (1996).

[40] G. Kresse and J. Furthmüller, Phys. Rev. B **54**, 11169 (1996).

[41] P.E. Blöchl, Phys. Rev. B **50**, 17953 (1994).

[42] G. Kresse and D. Joubert, Phys. Rev. B **59**, 1758 (1999).

[43] J.P. Perdew, K. Burke, and M. Enzerhof, Phys. Rev. Lett. **77**, 3865 (1996).

[44] S. Grimme, J. Comput. Chem. **27**, 1787 (2006).

[45] S. Grimme, Wiley Interdiscip. Rev. Comput. Mol. Sci. **1**, 211 (2011).

[46] X. Duan, O. Warschkow, A. Soon, B. Delley, and C. Stampfl, Phys. Rev. B **81**, 075430 (2010).

[47] L. Bengtsson, Phys. Rev. B **59**, 12301 (1999).

[48] H.J.H. Monkhorst and J.D.J. Pack, Phys. Rev. B **13**, 5188 (1976).

[49] G. Henkelman, B.P. Uberuaga, and H. Jónsson, J. Chem. Phys. **113**, 9901 (2000).

[50] J. Tersoff and D.R.D. Hamann, Phys. Rev. B **31**, 805 (1985).

[51] Http://Www.P4vasp.At/

[52] J. Heyd, G.E. Scuseria, and M. Ernzerhof, J. Chem. Phys. **118**, 8207 (2003).

[53] D. Sekiba, T. Inokuchi, and Y. Wakimoto, Surf. Sci. **470**, 43 (2000).

[54] G. Binnig, K. Frank, H. Fuchs, N. Garcia, B. Reihl, H. Rohrer, F. Salvan, and A. Williams, Phys. Rev. Lett. **55**, 991 (1985).

[55] W.M. Haynes, D.R. Lide, and T.J. Bruno, CRC Handb. Chem. Physics, 97th Ed. 12 (2017).

[56] C.D. Ruggiero, T. Choi, and J.A. Gupta, Appl. Phys. Lett. **91**, 253106 (2007).

STM and DFT studies of CO$_2$ adsorption on O-Cu(100) surface

Steven J. Tjung[1,†], Qiang Zhang[2,†], Jacob J. Repicky[1], Simuck F. Yuk[2], Xiaowa Nie[3], Nancy M. Santagata[1], Aravind Asthagiri[2], Jay A. Gupta[1]*

[1]*Department of Physics, The Ohio State University, Columbus, OH 43210*

[2]*William G. Lowrie Department of Chemical and Biomolecular Engineering, The Ohio State University, Columbus, OH 43210*

[3] *School of Chemical Engineering, PSU-DUT Joint Center for Energy Research, State Key Laboratory of Fine Chemicals, Dalian University of Technology, Dalian 116024, P. R. China*

# Supplementary Information

**Comparison of PBE & HSE06-DFT**

Figure S1 shows a simulated STM image of the Cu(100)-O surface with the same setup (-400 mV bias, 0.001 $e$/Å$^3$ isosurface density) using PBE versus HSE06. Both STM images exhibit the ladder structure with bright contrasts corresponding to surface Cu atoms and dark contrasts corresponding to the missing row. Aside from the different contrast of the overall image, the major difference between the PBE and HSE06 images is that the PBE additionally shows gray contrasts between rungs, corresponding to the surface O atoms. Such additional contrasts make the ladder-like structure less clear using PBE and the HSE06 version more resemble the experimental image. The different contrast of the O atoms with PBE versus HSE06 for this bias setting can be understood by examining the LDOS for the surface O atom. Figure S2 shows that HSE06 shifts the LDOS of the O surface atom down in energy versus PBE. At a bias voltage of -400 mV this means that within HSE06 the O surface atom will contribute less to the VASP generated partial charge density. This results in the O surface atom registering as a dark contrast in the HSE06 STM image versus a gray contrast in the PBE STM image.

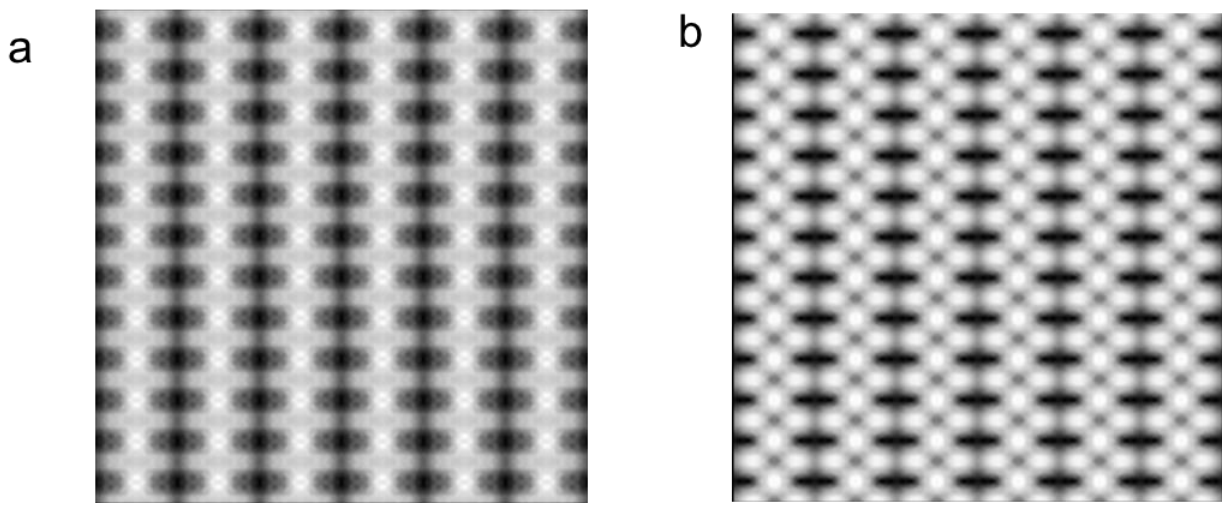

**Figure S1.** Simulated STM using (a) PBE and (b) HSE06 for O-Cu(100) surface with a voltage bias of -400 mV and an isosurface density of 0.001 $e$/Å$^3$.

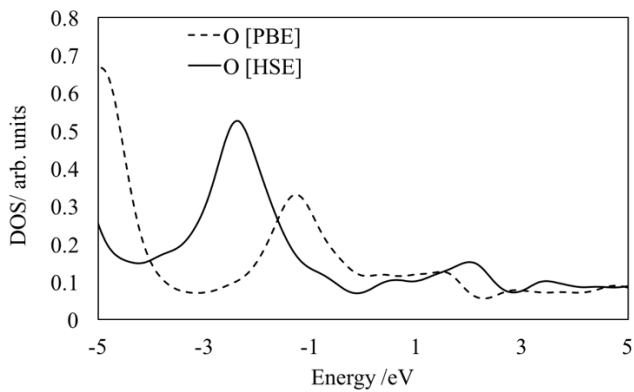

**Figure S2.** LDOS for the surface O atom on O-Cu(100) using PBE and HSE06 functionals.

**Motion of CO$_2$ molecules**

While most of the tip-induced motion of the CO$_2$ molecules was apparent as glitch lines in STM images, we also observed motion of the molecules without glitch lines. Figure S3 shows two successive STM images showing motion of a CO$_2$ molecule without glitch lines. We interpret these images not as an indication of intrinsic surface diffusion, but rather that the STM tip can perturb molecules > 1 nm away. In this case, molecules can be perturbed while the STM tip is scanning nearby regions of the image, and molecular motion will not appear as glitch lines.

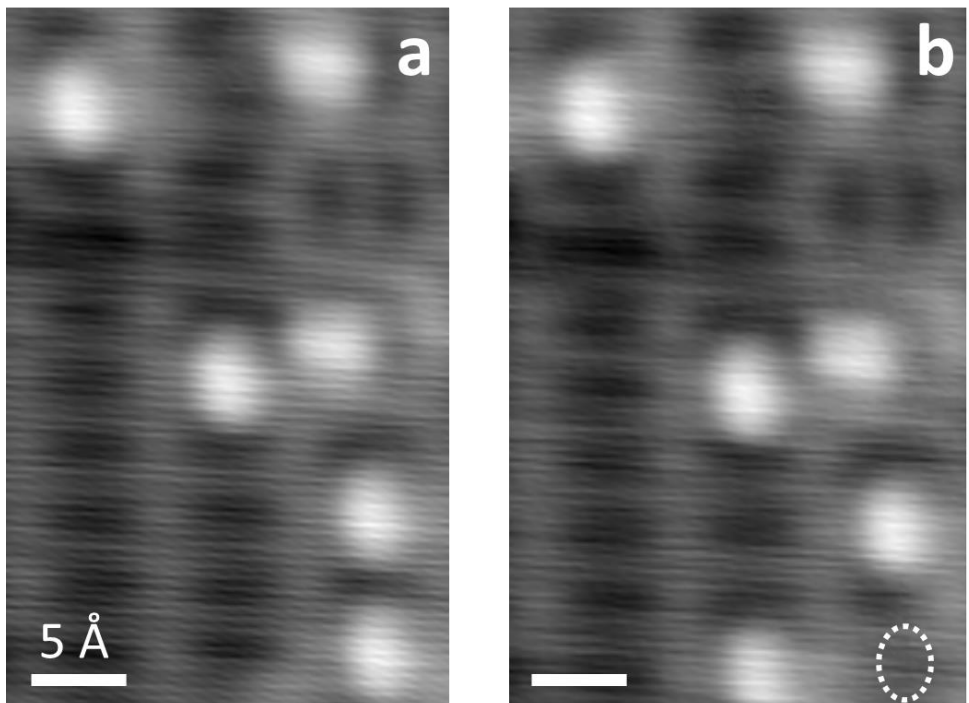

**Figure S3.** Successive STM images of tip-induced motion of the CO$_2$ molecules without glitch lines. The CO$_2$ molecule on the bottom right of figure (a) hopped across a CuO column as seen in (b). Previous location of the CO$_2$ molecule is marked with white dash oval in (b).

In support of this interpretation, we confirmed that the CO$_2$ molecules are stable on the Cu(100)-O surface at 5K in the absence of tip perturbation. Figure S4 shows STM images of the same nanoscale area taken 12 hours apart with the tip retracted ~ 300 nm between images. The pattern of the bright protrusions locating the CO$_2$ molecules is unchanged, indicating that no surface diffusion has taken place in this time period.

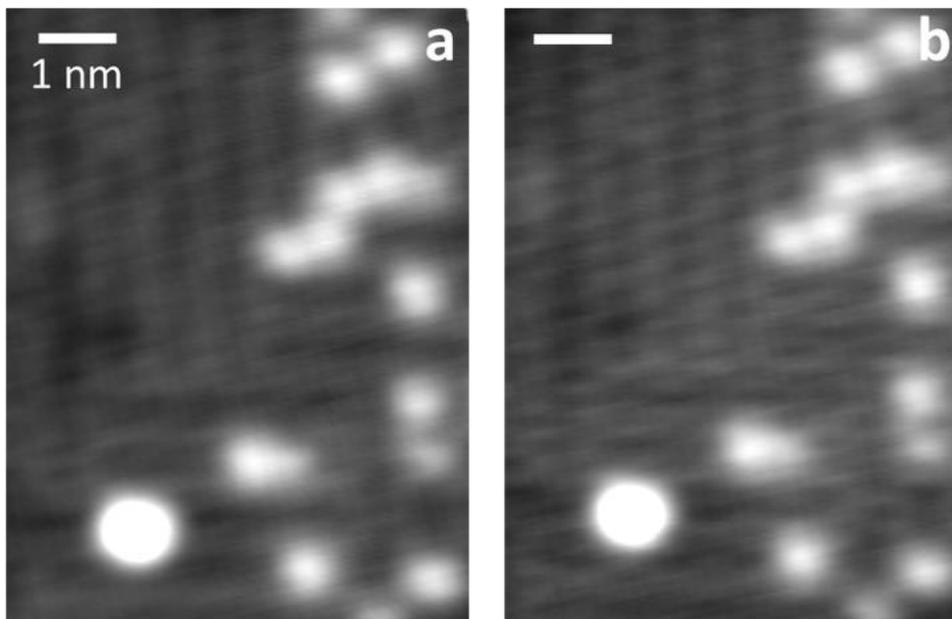

**Figure S4.** (a) STM image of $CO_2$ molecules on Cu(100)-O surface. (b) STM image of the same area in (a) taken 12 hours later showing no motion of the $CO_2$ molecules.

**Additional DFT calculations for CO2 adsorption**

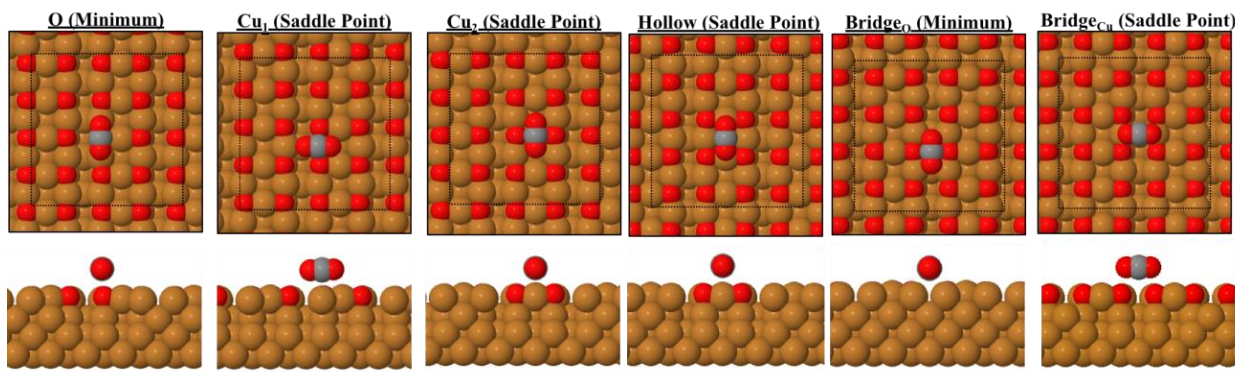

**Figure S5.** DFT PBE relaxed $CO_2$ adsorption configuration and binding energy for the six sites examined on Cu(100)-O.

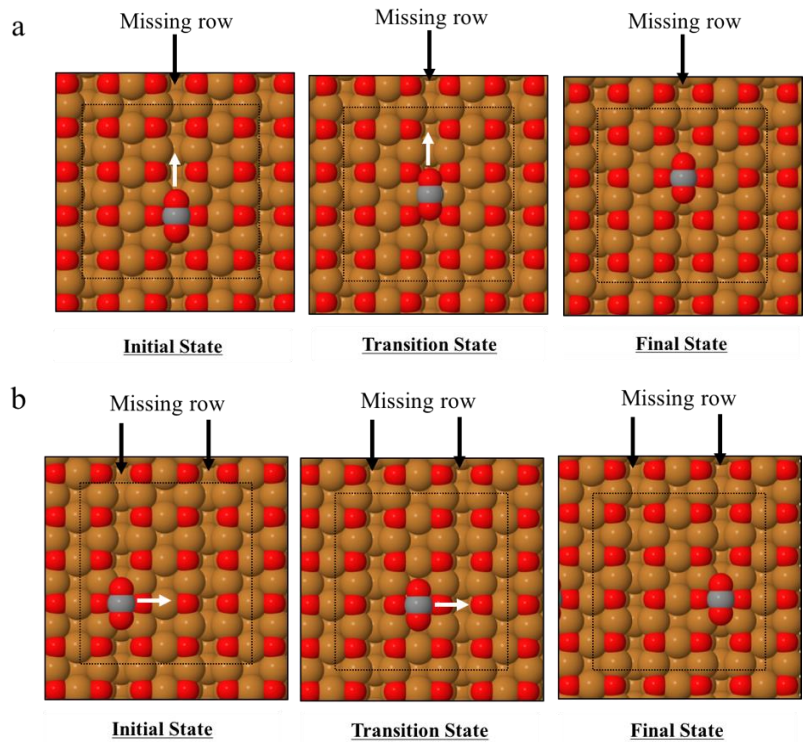

**Figure S6.** The diffusion path obtained from cNEB for $CO_2$ diffusion: (a) along the missing row, (b) across the missing row

## DFT-calculated vibrational frequencies

As is shown in Table S1, there is error between experimental gas phase $CO_2$ vibrational frequencies and PBE values. Such error is well known issue with DFT vibrational frequencies but the change in frequency upon adsorption should be relatively accurate. PBE shows that $CO_2$ adsorption on Cu(100)-O has a minor effect the vibrational frequencies (only by an average of a decrease of 10 cm$^{-1}$), consistent with the weak adsorption of $CO_2$ on this surface.

**Table S1.** Vibrational frequencies (cm$^{-1}$) for experimental $CO_2$, PBE isolated $CO_2$, PBE adsorbed $CO_2$

| Vibrational types | CO$_2$,exp (gas) | CO$_2$,PBE (gas) | ratio of CO$_2$,PBE/CO$_2$,exp. | CO$_2$,PBE (adsorbed) |
|---|---|---|---|---|
| **Antisymmetric stretch** | 2349[*] | 2367 | 1.0076 | 2357 |
| **Symmetric stretch** | 1333[*] | 1318 | 0.9888 | 1315 |
| **In plane bending** | 667[-] | 631 | 0.9464 | 619 |
| **Out of plane bending** | 667[-] | 631 | 0.9462 | 613 |
| **Surface related** | - | - | - | 86 |
| **Surface related** | - | - | - | 71 |
| **Surface related** | - | - | - | 62 |
| **Surface related** | - | - | - | 57 |
| **Surface related** | - | - | - | 35 |

[*] Shimanouchi, T., Tables of Molecular Vibrational Frequencies, Consolidated Volume 1, NSRDS NBS-39

[-] "Vibrational Intensities in Infrared and Raman Spectroscopy" WB Person, G Zerbi, ed. Elsevier, Amsterdam, 1982

6